\documentclass[a4paper,11pt]{article}
\usepackage{pos}

\usepackage{booktabs}
\usepackage{tabularx}
\usepackage{upgreek}
\usepackage{xspace}
\usepackage[per-mode=symbol]{siunitx}
\usepackage{amsmath}
\usepackage[colorinlistoftodos]{todonotes} 
\usepackage{MnSymbol}
\usepackage{subfig}

\newcommand{\BR}{\ensuremath{\mathscr{B}}\xspace}
\newcommand{\R}{\ensuremath{\mathscr{R}}\xspace}
\newcommand{\order}{\ensuremath{\mathscr{O}}}

\newcommand{\CL}{\SI{90}{\percent}\,CL\xspace}

\newcommand{\muegamma}{\ensuremath{\mu\to e\gamma}\xspace}
\newcommand{\muegammacharged}{\ensuremath{\mu^+\to e^+\gamma}\xspace}
\newcommand{\mueee}{\ensuremath{\mu\to eee}\xspace}
\newcommand{\mueeecharged}{\ensuremath{\mu^+\to e^+e^-e^+}\xspace}
\newcommand{\muNeN}{\ensuremath{\mu N\to eN}\xspace}
\newcommand{\muNeNcharged}{\ensuremath{\mu^- N\to e^-N}\xspace}

\newcommand{\mueXcharged}{\ensuremath{\mu^+\to e^+X}\xspace}

\newcommand{\muegammaXcharged}{\ensuremath{\mu^+\to e^+\gamma X}\xspace}

\newcommand{\muenunucharged}{\ensuremath{\mu^+\to e^+\overline{\nu}_\mu\nu_e}\xspace}

\newcommand{\mueeenunucharged}{\ensuremath{\mu^+\to e^+e^-e^+\overline{\nu}_\mu\nu_e}\xspace}

\newcommand{\muegammanunucharged}{\ensuremath{\mu^+\to e^+\gamma\overline{\nu}_\mu\nu_e}\xspace}

\newcommand{\muNeNnunucharged}{\ensuremath{\mu^- N\to e^-\nu_\mu\overline{\nu}_eN}\xspace}

\DeclareSIUnit{\muons}{\mu}
\DeclareSIUnit\permille{\text{\textperthousand}}
\DeclareSIUnit{\ifb}{\femto\barn^{-1}}
\DeclareSIUnit{\iab}{\atto\barn^{-1}}

\title{A Review of $\mu\to eee$, $\mu\to e\gamma$ and $\mu N\to eN$ Conversion}

\author*[a]{Ann-Kathrin Perrevoort}

\onbehalf{on behalf of the Mu3e Collaboration}

\affiliation[a]{Karlsruhe Institute of Technology,
  Institute of Experimental Particle Physics,\\
  Hermann-von-Helmholtz-Platz 1, Eggenstein-Leopoldshafen, Germany}

\emailAdd{ann-kathrin.perrevoort@kit.edu}

\abstract{The observation of lepton flavour violation (LFV) in interactions
  involving charged leptons would be an unambiguous sign of physics beyond the
  Standard Model of particle physics. Given that muons can be produced at high
  intensities, searches for LFV with muons are particularly sensitive.
  
In a global initiative, ongoing and upcoming experiments are aiming to
discover physics beyond the Standard Model in the three golden muon LFV
channels: $\mu\to e\gamma$, $\mu\to eee$ and $\mu$-to-$e$ conversion on
nuclei. With innovative detector concepts and new muon beam lines, these
experiments will be able to investigate muon LFV in the coming
years with sensitivities improved by up to four orders of magnitude compared
to past searches. 

The current status of muon LFV searches is discussed
and the ongoing MEG II and DeeMe experiments as well as the upcoming Mu2e,
COMET and Mu3e experiments are presented.
}

\FullConference{21st Conference on Flavor Physics and CP Violation (FPCP 2023)\\
 29 May - 2 June 2023\\
 Lyon, France\\}

\renewcommand{\logo}{\relax}

\begin{document}
\maketitle

\section{Searches for Charged Lepton Flavour Violation with Muons}

In the original formulation of the Standard Model (SM) of particle physics,
lepton flavour is a conserved quantity, however, only due to an accidental
symmetry.
Hence, lepton flavour violation (LFV) is a common signature in numerous models
of physics beyond the SM (BSM)~\cite{Calibbi:2017uvl}.
Moreover, with the observation of neutrino oscillations it became evident that
LFV processes in the neutral lepton sector occur in
nature~\cite{Fukuda:1998mi, Ahmad:2001an, Eguchi:2002dm}. \\ 
LFV processes in the charged lepton sector, however, have eluded observation
so far.
If mediated solely via neutrino mixing, charged LFV (cLFV) interactions would
be suppressed to tens of orders of magnitude below the sensitivity of current
and near future experiments.
For the cLFV process \muegamma, for example, branching ratios $\BR$ below
$\num{e-54}$ are predicted~\cite{Petcov:1976ff, Bilenky:1977du,
  deGouvea:2013zba}.
Consequently, the observation of cLFV would be an unambiguous sign of physics
beyond the SM and beyond neutrino mixing.

There is a wide range of cLFV searches performed e.g.\ at general purpose and
$B$-physics experiments at colliders as well as specialised kaon and muon
physics experiments. \\
Among all types of cLFV searches, searches for $\mu$-to-$e$ transitions
yield the strongest limits today.
This is because muons can be effectively produced at high rates and decay to
comparably clean final states which include only electrons, neutrinos and
photons.
Typically, dedicated experiments are operated at high-intensity muon sources
which perform background-free searches for a selected $\mu$-to-$e$ LFV
channel.
These experiments have to meet not only the demands with regard to precision
for an effective background suppression, but also have to cope with challenges
arising from high muon decay rates in terms of data processing and background
from accidental combinations.
In addition, the muons are typically stopped in these experiments which
results in decay electrons with momenta of $\order(\SI{10}{\MeV})$ for which
the momentum resolution is significantly deteriorated by multiple Coulomb
scattering.
The experiments therefore rely on detectors with minimised material amount in
the active detector volume.

There is a global initiative to push the sensitivity of the so-called three
golden channels \muegammacharged, \mueeecharged and $\mu$-to-$e$
conversion on nuclei \muNeNcharged by orders of magnitude in the coming years:
the ongoing MEG~II and DeeMe experiments as well as the upcoming COMET, Mu2e
and Mu3e experiments.
These experiments are presented in the following.
In addition, the complementarity of the three golden channels as well as
further options for BSM searches at muon cLFV experiments are discussed.

\section{Searches for \muegamma}

\begin{figure}
  \centering
  \subfloat[Longitudinal and transverse view of the MEG
  experiment~\cite{MEG:2016leq}. ]{\includegraphics[height=0.19\textheight]{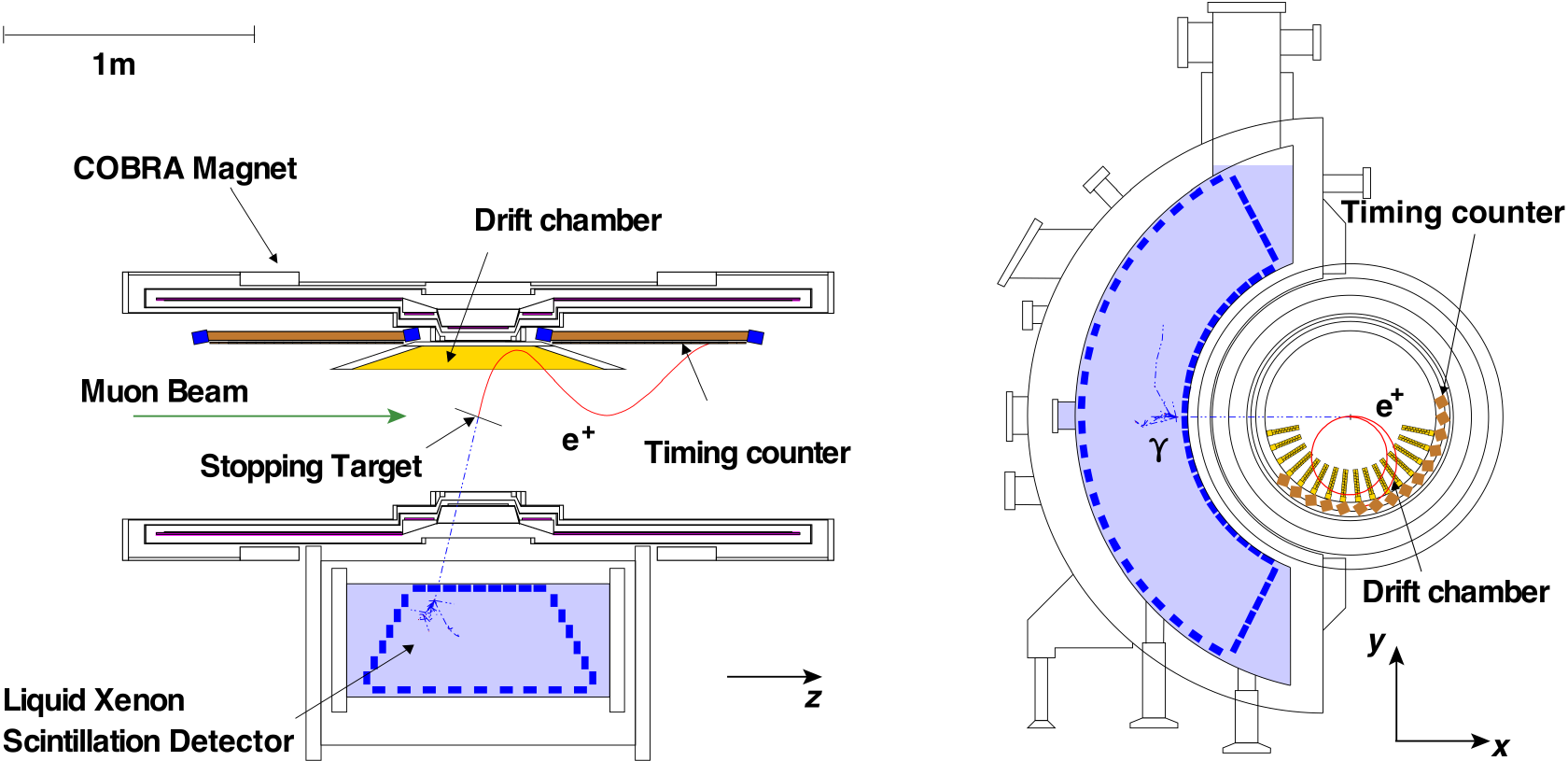}\label{fig:MEGIdet}}\hfill
  \subfloat[3D view of the MEG~II
  experiment~\cite{MEGII:2018kmf}. ]{\includegraphics[height=0.19\textheight]{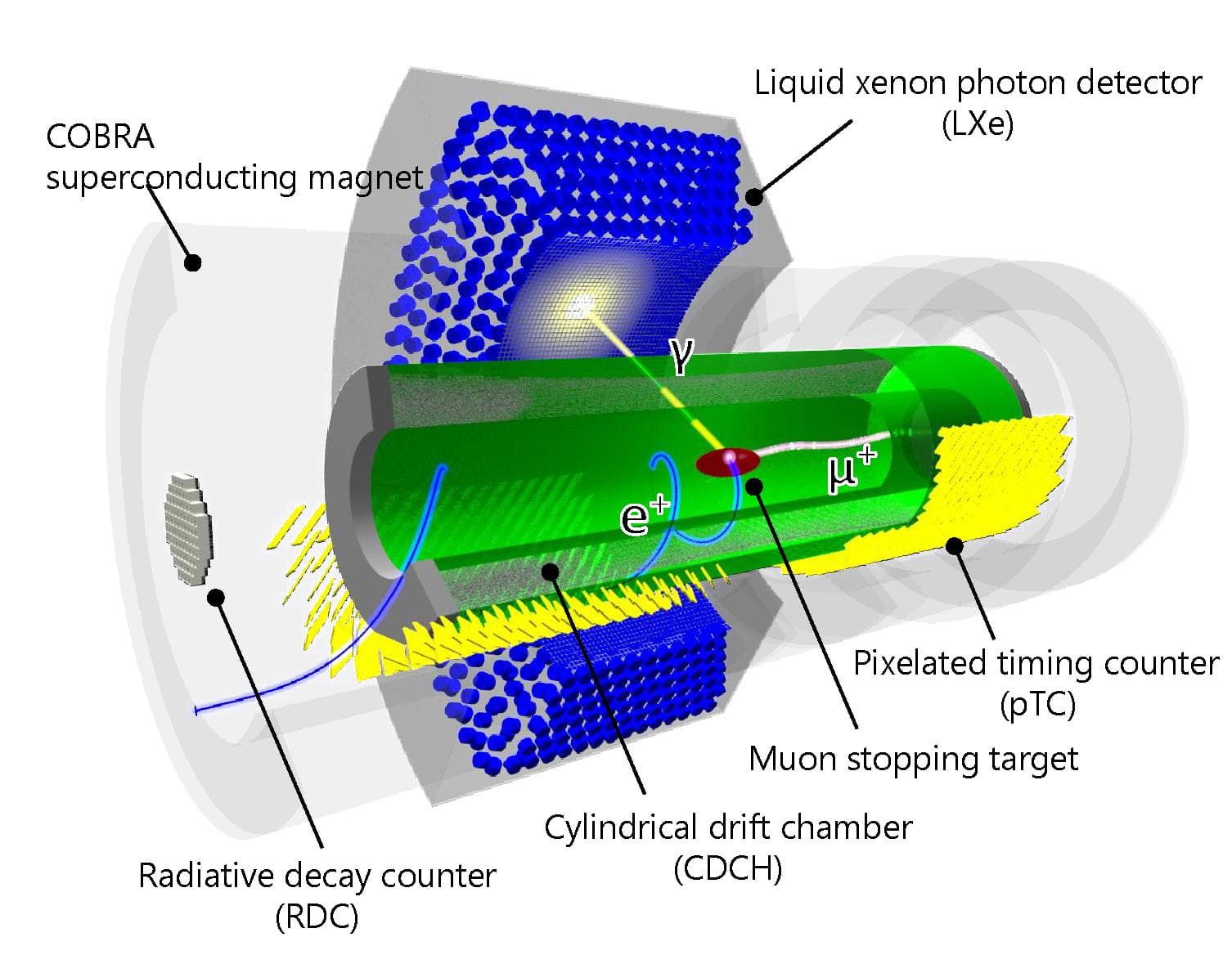}\label{fig:MEGIIdet}}
  \caption{Schematics of the MEG and MEG~II experiments searching for
    \muegamma~\cite{MEG:2016leq, MEGII:2018kmf}. }
  \label{fig:MEGdet}
\end{figure}

\begin{figure}
  \centering
  \subfloat[Event distribution in the $(E_e, E_\gamma)$ plane. ]{\includegraphics[width=0.45\textwidth]{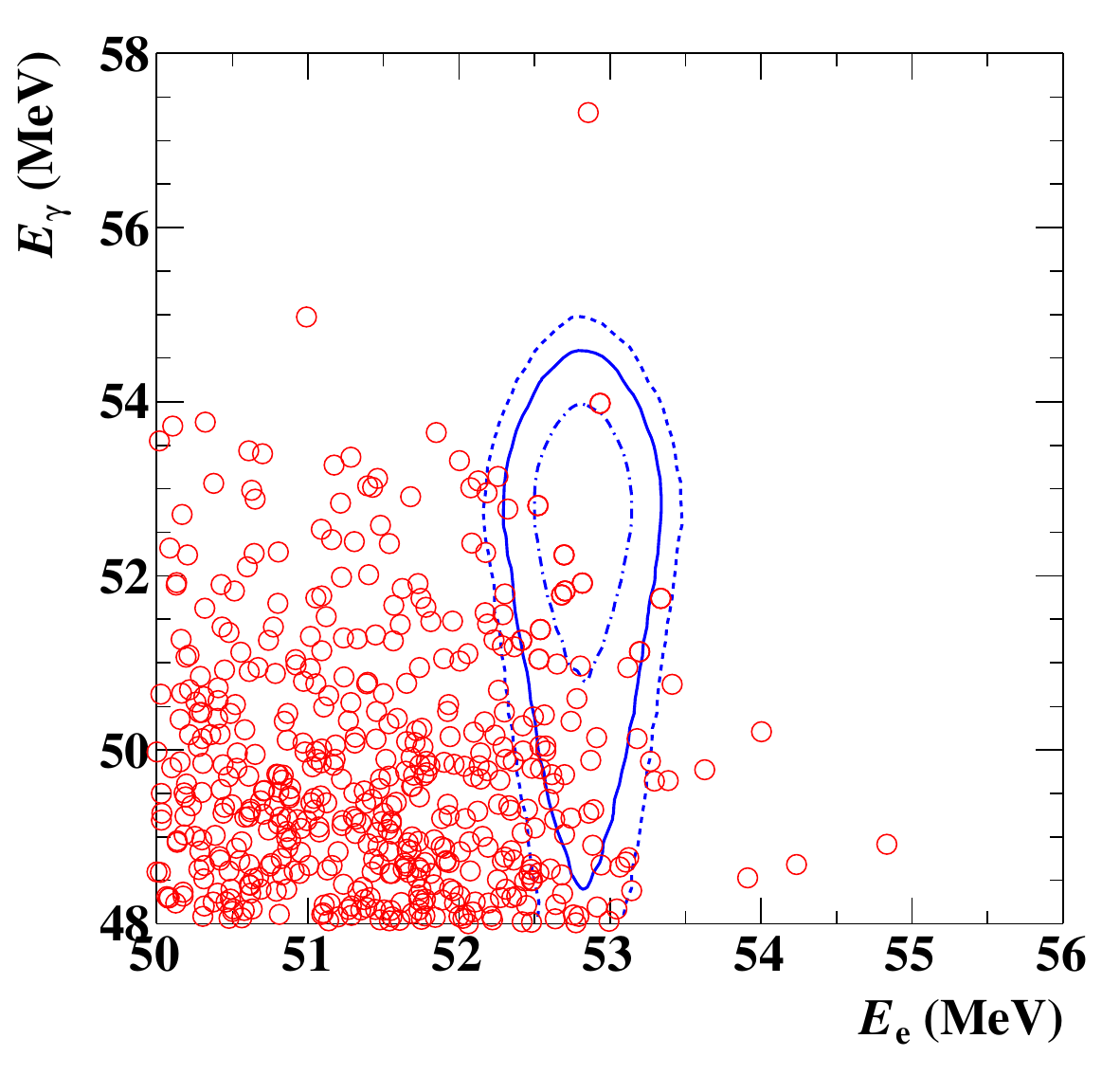}}\hfill
  \subfloat[Event distribution in the $(\cos\Theta_{e\gamma}, t_{e\gamma})$ plane. ]{\includegraphics[width=0.45\textwidth]{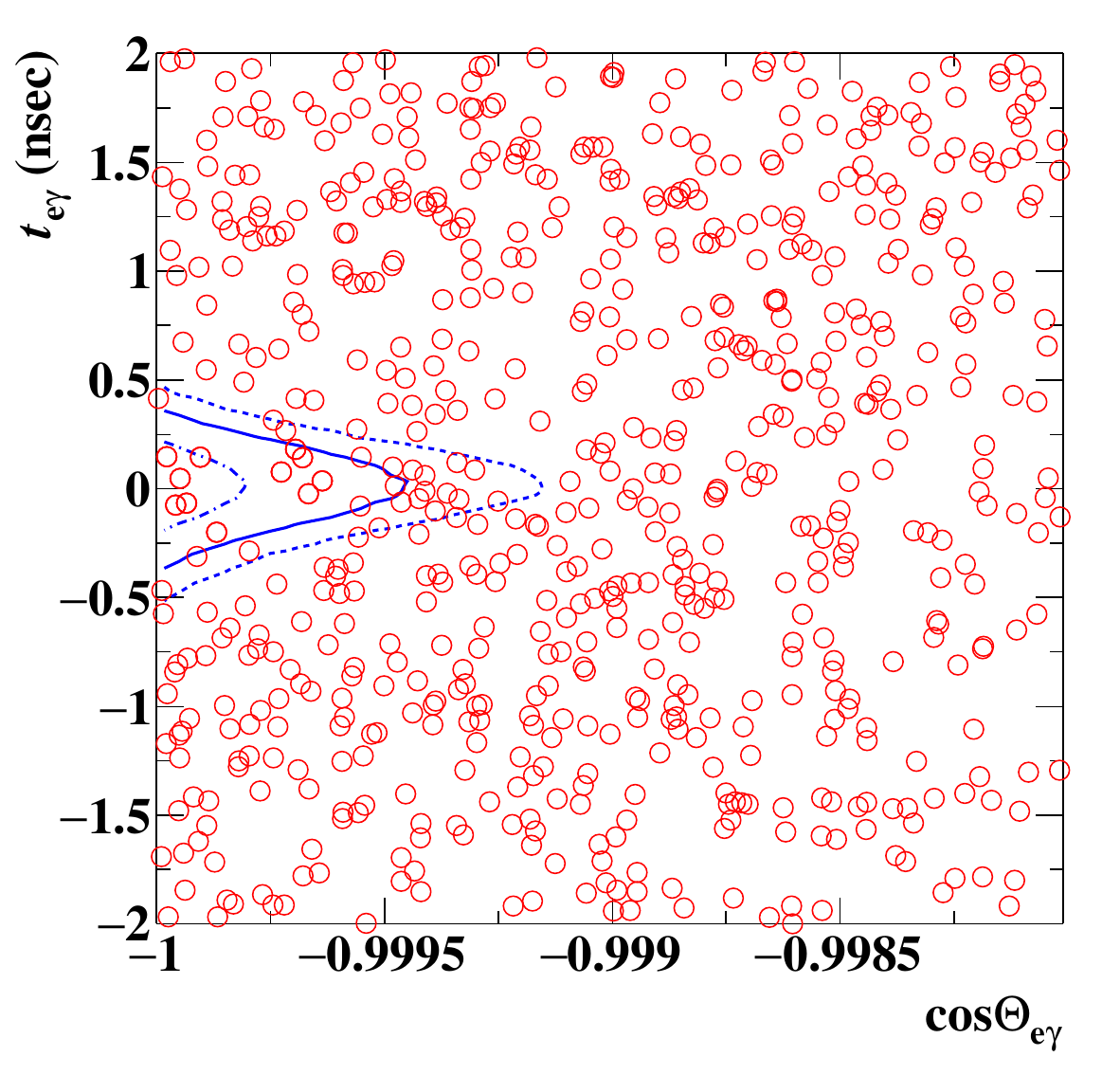}}
  \caption{Distributions of observed events (red circles) and contours
    ($1\sigma, 1.64\sigma, 2\sigma$) of the signal probability density
    function (blue dashed-dotted, solid and dotted lines).
    $\Theta_{e\gamma}$ denotes the relative stereo angle of the positron and
    photon~\cite{MEG:2016leq}. }
  \label{fig:MEGresult}
\end{figure}

The latest and currently most stringent limits on cLFV processes have been set
by the MEG collaboration on the decay \muegammacharged~\cite{MEG:2016leq}.\\
In the rest frame of the decaying muon, the signature of the signal decay is
a mono-energetic positron and photon emitted back-to-back with an energy of
half the muon rest mass.
Background events stem from the radiative muon decay \muegammanunucharged in
which the neutrinos are not detected, as well as from accidental coincidences
of positrons and photons from different origins such as SM muon decays,
Bremsstrahlung or positron annihilation.
The first type of background is suppressed by precise measurements of the
energy and direction of the positron and photon, and the latter type by an
accurate measurement of their relative timing as well as by operating with a
continuous muon beam. \\
The MEG experiment is depicted in Fig.~\ref{fig:MEGIdet}.
The experiment was operated from 2009 to 2013 at the $\pi$E5 channel at the
Paul Scherrer Institute (PSI).
The incoming continuous $\mu^+$ beam is stopped on a slanted target in the
centre of the detector at a stopping rate of \SI{3e7}{\muons\per\second}.
The trajectory and momentum of positrons is measured with a low-material
drift chamber system placed in the COnstant Bending RAdius magnet with a
gradient field between \SI{1.27}{\tesla} to \SI{0.49}{\tesla}.
The COBRA magnet has the advantage that the bending radius of the $e^+$
trajectory depends only weakly on the emission angle and that $e^+$ are
quickly swept out of the detector even at low longitudinal momenta.
An additional timing counter built from scintillating bars with
photo-multiplier (PMT) readout is installed for precise measurements of the
impact time and position of $e^+$.
The energy and timing of the photon is measured with a Liquid Xenon
scintillation detector which is read out with PMTs.
The direction of the photon is inferred from the interaction vertex in the LXe
and the reconstructed intersection point of the trajectory of a matching
positron with the target surface.
The data acquisition system is triggered by information from the LXe and
timing counter detectors on the photon energy and the relative direction and
timing of the positron and photon.\\
The MEG collaboration has performed a blinded, maximum likelihood analysis on
a data set of $\num{7.5e14}$ $\mu^+$ decays.
As observables, the positron and photon energy $E_{e}$ and $E_\gamma$, as well
as the relative time $t_{e\gamma}$ and the relative azimuthal and polar angles
$\theta_{e\gamma}$ and $\phi_{e\gamma}$ of the positron and photon were used
to distinguish signal from background events.
Event distributions are shown in Fig.~\ref{fig:MEGresult}.
No significant excess with respect to the expected background was found and an
upper limit on the branching ratio of \muegammacharged was set at
$\BR(\muegammacharged)<\num{4.2e-13}$ at \SI{90}{\percent}\,confidence level
(CL)~\cite{MEG:2016leq}.

The MEG experiment underwent a substantial upgrade---MEG~II---in order to
boost the sensitivity of the \muegammacharged search by increasing the rate
capability and by improving the resolution of the detector (see
Fig.~\ref{fig:MEGIIdet}). \\
The drift chamber has been replaced by a single-volume, cylindrical drift
chamber with high wire density and the timing counter by a pixelated timing
counter built from scintillating tiles with Silicon Photon Multiplier (SiPM)
readout.
The PMTs on the entrance surface of the LXe detector were replaced by SiPMs
for higher granularity.
In addition, a new detector, the radiative decay counter, was installed
downstream of the target in order to veto positrons from background
processes.
The radiative decay counter consists of scintillating crystals and plastic
scintillators for both a precise energy and timing measurement.
In this way, MEG~II is capable to operate at stopping rates of
\SI{7e7}{\muons\per\second} making full use of the muon beam rates available
at PSI. \\
The MEG~II experiment is taking physics data since 2021.
It is expected to achieve a sensitivity to $\BR(\muegammacharged)$ down to
$\num{6e-14}$ at \CL~\cite{MEGII:2018kmf}.

\section{Searches for \mueee}

Unlike the other golden muon LFV channels, the \mueeecharged decay is not
characterised by mono-energetic particles in the final state.
Background stems on the one hand from the SM \mueeenunucharged decay which can
be distinguished from signal decays only by the missing momentum from the
undetected neutrinos.
On the other hand, background events are generated by accidental combinations
of electrons and positrons from various origins such as the dominating
\muenunucharged decay, Bhabha scattering and photon conversion as well as from
misreconstruction.
The suppression of the first type of background requires an excellent momentum
resolution in the tracking of electrons and positrons, and the suppression of
the latter type accurate vertex reconstruction and timing measurement as well
as a low material detector operated at a continuous muon beam. \\
Current limits on \mueeecharged stem from the SINDRUM experiment which
operated at PSI~\cite{SINDRUM:1987nra}.
The SINDRUM experiment studied $\mu^+$ decays at rest with a spectrometer
built from concentric multiwire proportional chambers (MWPC) and a hodoscope
made of scintillation counters placed in a \SI{0.33}{\tesla} solenoidal
field.
The collaboration found no events in the signal-sensitive region and set an
upper limit on the branching ratio at $\BR(\mueeecharged)<\num{1.0e-12}$ at
\CL.

\begin{figure}
  \centering
  \subfloat[Longitudinal view.]{\includegraphics[height=0.165\textheight]{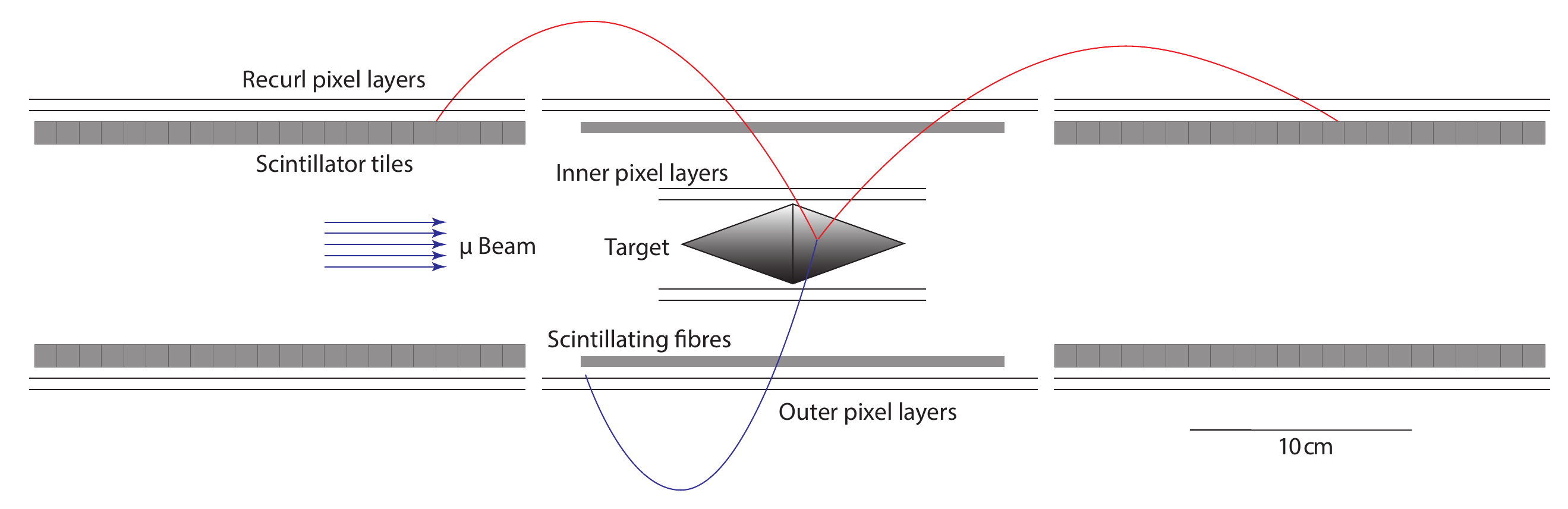}}\hfill
  \subfloat[Transverse view.]{\includegraphics[height=0.165\textheight]{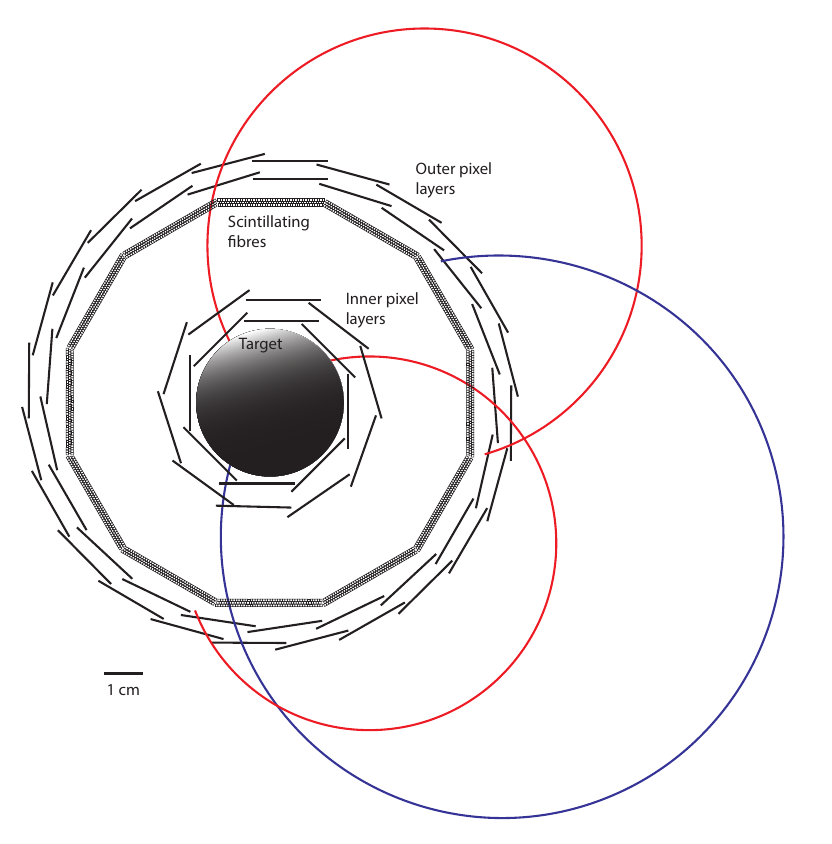}}
  \caption{Schematic of the Mu3e detector in phase~I~\cite{Mu3eTDR}.
    The recurl stations are placed upstream and downstream of the central
    detector station.
    A \mueeecharged signal decay is shown with positron trajectories in red
    and the electron trajectory in blue.}
  \label{fig:Mu3edet}
\end{figure}

\begin{figure}
  \centering
  \subfloat[Distribution of simulated signal and background events in the
  invariant $e^+e^-e^+$ mass 
  from $\num{e15}$ $\mu^+$ decays at a stopping rate of
  \SI{e8}{\muons\per\second}. ]{\includegraphics[height=0.23\textheight]{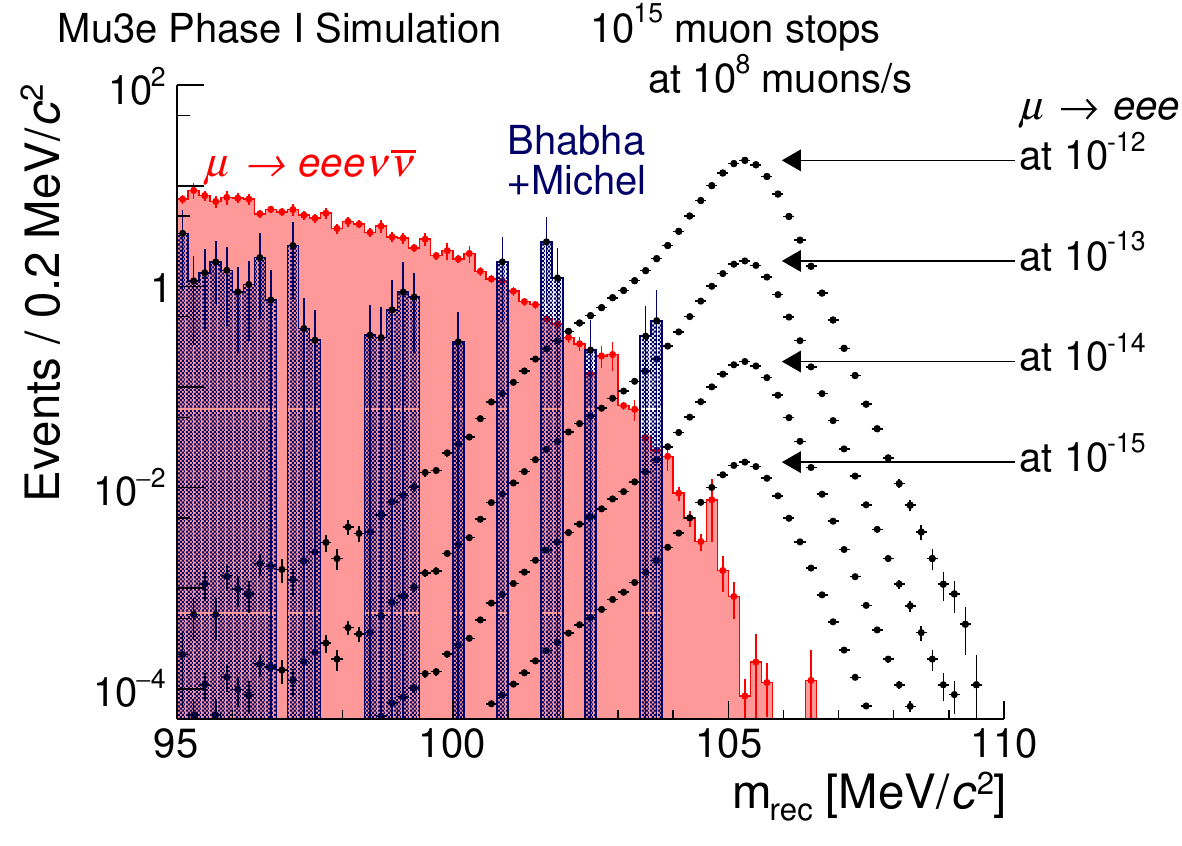}\label{fig:Mu3eSimInvMass}}\hfill
  \subfloat[Expected sensitivity 
  on $\BR(\mueeecharged)$ in dependence of the running days at a stopping rate
  of \SI{e8}{\muons\per\second}. ]{\includegraphics[height=0.23\textheight]{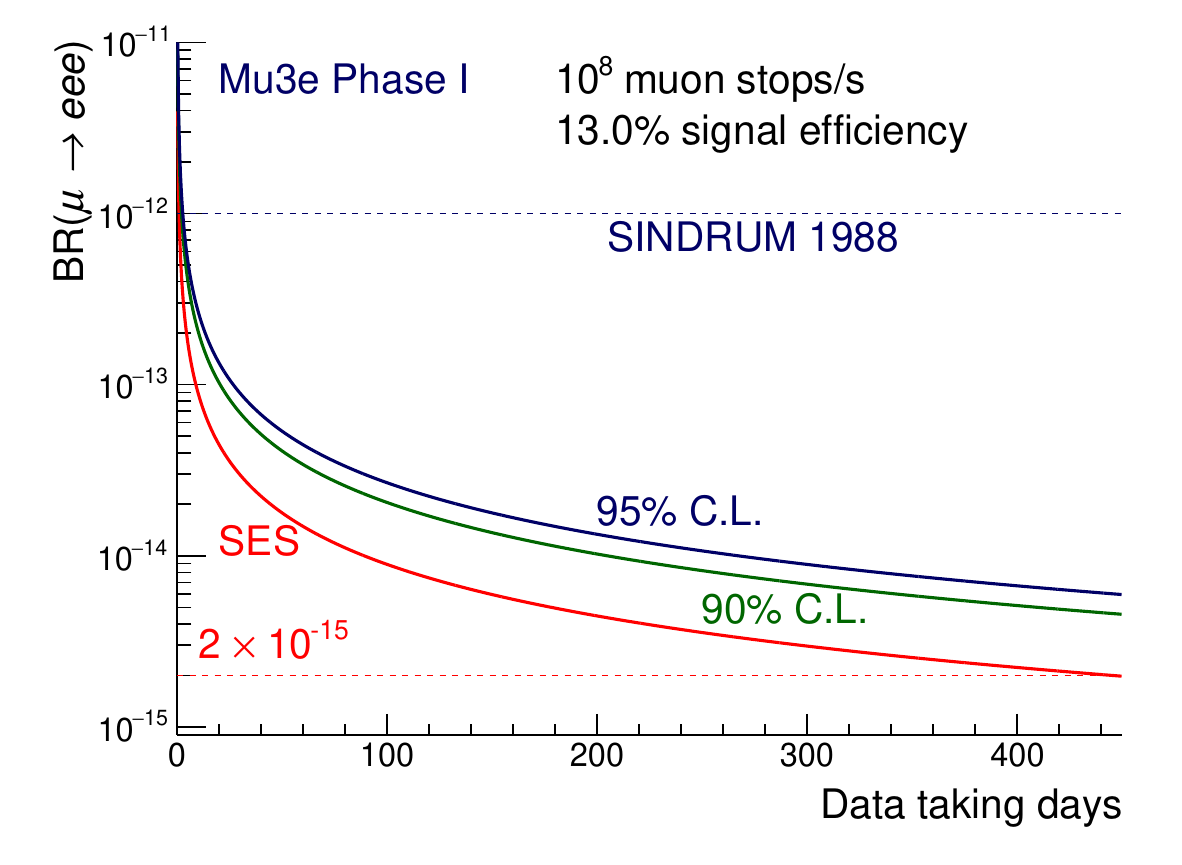}\label{fig:Mu3eSimSensitivity}}
  \caption{Simulated performance of the Mu3e experiment in phase~I~\cite{Mu3eTDR}.}
  \label{fig:Mu3eSim}
\end{figure}

More than three decades later, the upcoming Mu3e experiment aims to repeat the
search for \mueeecharged with a sensitivity improved by four orders of
magnitude compared to SINDRUM. \\
The Mu3e experiment will be conducted in two phases.
The phase~I experiment is depicted in Fig.~\ref{fig:Mu3edet}.
The incoming, continuous $\mu^+$ beam is stopped on a hollow, double-cone
target in the centre of the experiment at a stopping rate of
\SI{e8}{\muons\per\second}. 
The detector is a low-material tracking detector built from novel, ultra-thin
Silicon pixel sensors in High Voltage Monolithic Active Pixel Sensor
technology~\cite{Peric:2007zz} for precise tracking of electrons and
positrons.
The detector is placed in a \SI{1}{\tesla} solenoidal magnetic field with a
magnet bore large enough to enable the electrons and positrons to return---or
\emph{recurl}---to the detector and be measured again.
For recurling particles which have performed a half turn in-between
measurements, scattering effects that deteriorate the momentum resolution
cancel to first order. 
The detector geometry is therefore optimised for a high acceptance of
recurling particles by installing recurl stations upstream and downstream of
the central detector station. 
Additional scintillating detectors, i.e.\,scintillating fibres in the central
station and scintillating tiles in the recurl stations, are installed for
accurate timing measurements. \\
The Mu3e experiment operates without a traditional hardware trigger.
Instead, all sub-detectors continuously stream data to the event filter farm
where online track reconstruction and vertex finding are performed on Graphics
Processing Units.
Events with \mueeecharged candidates are selected and stored for elaborate
offline reconstruction at optimum resolution and analysis. \\
Background events will be suppressed by selections on the relative timing of
the electron and positrons, the quality and position of the reconstructed
vertex and kinematic observables such as the invariant mass.
A distribution of simulated signal and background events from $\num{e15}$
$\mu^+$ decays is shown in Fig.~\ref{fig:Mu3eSimInvMass}.
The phase~I Mu3e experiment has an expected sensitivity on
$\BR(\mueeecharged)$ of a few $\num{e-15}$ at \CL depending on the running
time (see Fig.~\ref{fig:Mu3eSimSensitivity}). 
The Mu3e phase~I experiment is currently under construction.
First physics data runs are planned in 2025. 

The Mu3e phase~II experiment will be operated at muon stopping rates of
\SI{2e9}{\muons\per\second} at the upcoming High-Intensity Muon
Beams (HIMB) project at PSI~\cite{Aiba:2021bxe}.
Paired with an increased acceptance for recurling tracks and improved vertex
and timing resolution, sensitivities of $\num{e-16}$ at \CL are expected. 
The optimum design and detector technologies are currently under
investigation.

\section{Searches for \muNeN}

\begin{figure}
  \centering
  \subfloat[Schematic of the DeeMe experiment~\cite{DeeMeOnline}. ]{\includegraphics[height=0.16\textheight]{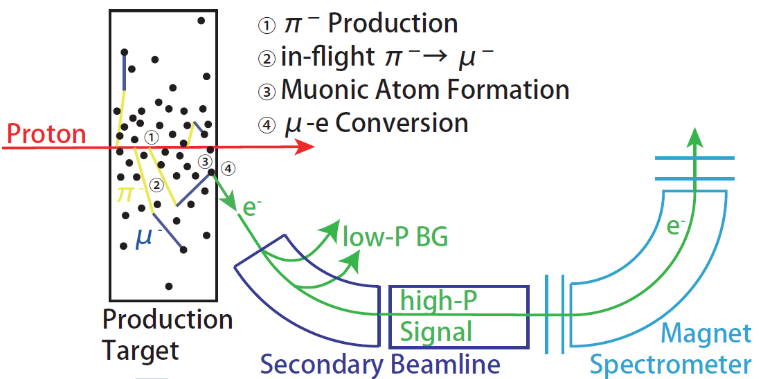}\label{fig:DeeMe}}\hfill
  \subfloat[Beam timing at the Mu2e experiment~\cite{BernsteinMu2e2019}. \emph{POT}
  stands for \emph{proton-on-target}. ]{\includegraphics[height=0.13\textheight]{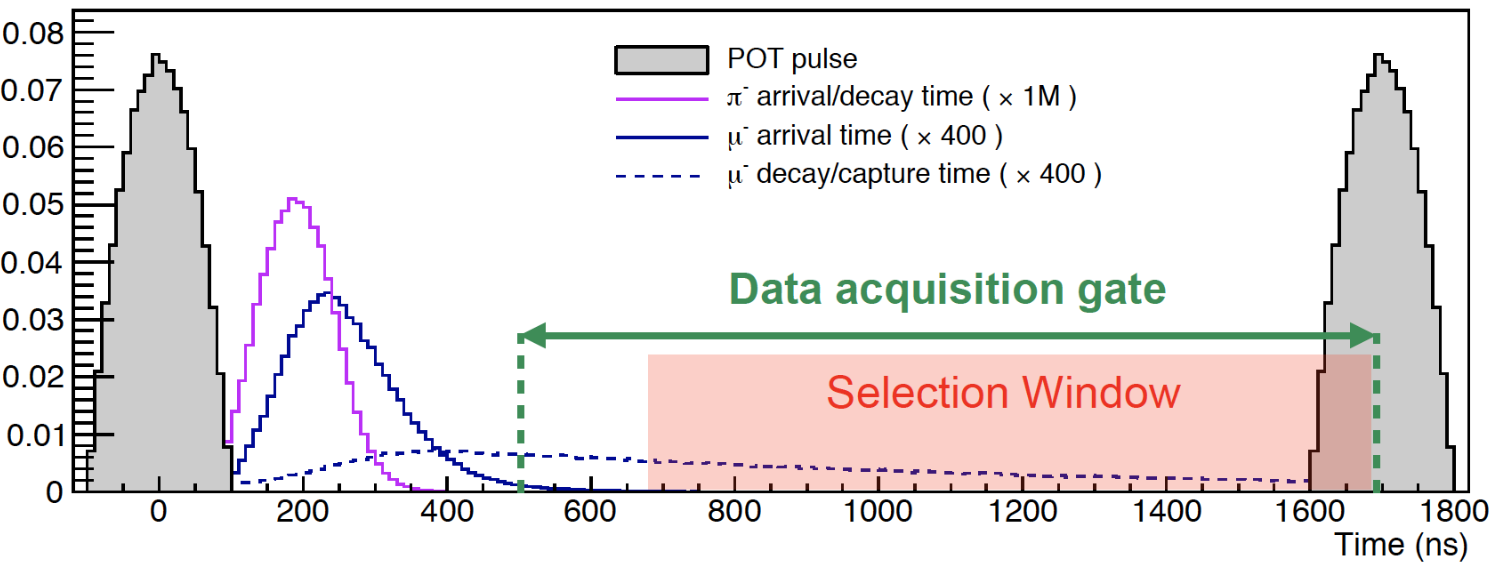}\label{fig:muNeNpulse}}
  \caption{Principle of \muNeNcharged searches. }
  \label{fig:muNeN}
\end{figure}

Another channel to search for LFV with muons are $\mu$-to-$e$ conversions
in muonic atoms: \muNeNcharged.\\
The measurement principle is shown in Fig.~\ref{fig:DeeMe}.
As in the previously discussed experiments, the $\mu^-$ beam is produced from
a proton beam hitting a production target and the decay of the resulting pions
and other hadrons to muons.
The muons are then stopped on a stopping target where muonic atoms are formed.
The \muNeNcharged decay is characterised by a mono-energetic electron which
energy is determined by the muon rest mass, the atomic binding energy of the
muon and the atomic recoil energy. \\
Background stems from muon decays in orbit $\muNeNnunucharged$ or can be
beam-induced for example from pions, electrons and anti-protons as well as
muon decays in flight.
In addition, cosmic rays can generate background events via decay, scattering
of interactions with material in the experiment.
The first type of background is suppressed by an excellent energy or momentum
measurement while for the last one dedicated cosmic ray vetoes are installed.
Beam-induced background decays faster than muonic atoms and can therefore be
reduced by the usage of a pulsed proton beam and by starting data recording
only several \SI{100}{\nano\second} after the proton pulse (see
Fig.~\ref{fig:muNeNpulse}). 
Target materials are typically chosen with the requirement of long lifetimes
for the muonic atom. 

\begin{figure}
  \centering
  \includegraphics[width=0.45\textwidth]{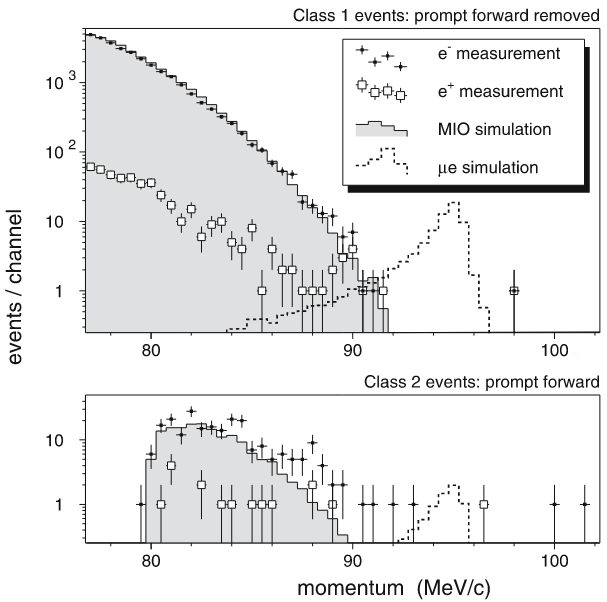}
  \caption{Distributions of the momentum of observed electrons and positrons
    as well as signal and background predictions in the SINDRUM~II
    experiment.
    Beam-induced background is correlated with the proton pulse on the
    production target and thus appears promptly with the arrival of muons on
    the stopping target and tends to be forward-directed. 
    \emph{MIO} stands for \emph{muon decay in orbit}. 
    ~\cite{SINDRUMII:2006dvw}}
  \label{fig:SINDRUMIIresult}
\end{figure}

The current most stringent limits on \muNeNcharged stem from the SINDRUM~II
experiment at PSI.
The SINDRUM~II experiment used a Gold foil as stopping target and measured the
electron trajectories with a combination of a drift chamber and a scintillator
and Cerenkov hodoscope placed in a solenoidal magnetic field.
Proton pulses were generated every \SI{19.75}{\nano\second}.
SINDRUM~II collected a total of \SI{4.37\pm0.32e13}{\mu^-} stopped on target.  
The momentum distributions of the observed events are shown in
Fig.~\ref{fig:SINDRUMIIresult}.
No significant excess above the expected background was found and an upper
limit on the conversion rate was set at $\R(\mu^-\text{Au}\to
e^-\text{Au})<\num{7.0e-13}$ at \CL~\cite{SINDRUMII:2006dvw}.

Three experiments will repeat the search for \muNeNcharged in the near future:
DeeMe~\cite{Natori:2014yba} and COMET~\cite{COMET:2018auw} at J-PARC and
Mu2e~\cite{Mu2e:2014fns} at Fermilab.

The DeeMe experiment will search for $\mu$-to-$e$ conversion on Carbon.
A schematic of the experiment in shown in Fig.~\ref{fig:DeeMe}.
The \SI{3}{\giga\eV} Rapid Cycling Synchrotron delivers proton double pulses
at a frequency of \SI{25}{\hertz} allowing to have \SI{10}{\micro\second} long
data taking intervals starting around \SI{300}{\nano\second} after the second
pulse. 
The Carbon target serves as both muon production and muon stopping target.
With a secondary beamline, electrons from signal processes are separated from
electrons from background processes which tend to have lower momenta.
Positively charged particles are removed, as well.
The electrons momenta are then measured in a magnet spectrometer consisting of
a \SI{0.4}{\tesla} dipole magnet bending the beam by \SI{70}{\degree} and two
times two thin MWPCs upstream and downstream of the magnet. 
The high voltage at the MWPCs is switched to a high gas gain mode only during
the data acquisition windows for detector protection in intervals of high
prompt background. \\
The DeeMe collaboration has performed commissioning runs and is getting ready
to take physics data.
A sensitivity to the conversion rate on Carbon of $\R_{\mu e}(\mu^-\text{C}\to
e^-\text{C})\lesssim\num{2e-13}$ at \CL is expected. 

\begin{figure}
  \centering
  \subfloat[The COMET phase~I experiment. ]{\includegraphics[height=0.28\textheight]{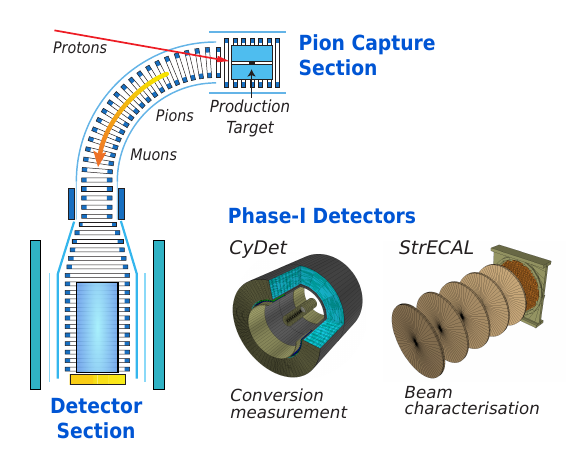}\label{fig:COMETIdet}}\hfill
  \subfloat[The COMET phase~II experiment. ]{\includegraphics[height=0.28\textheight]{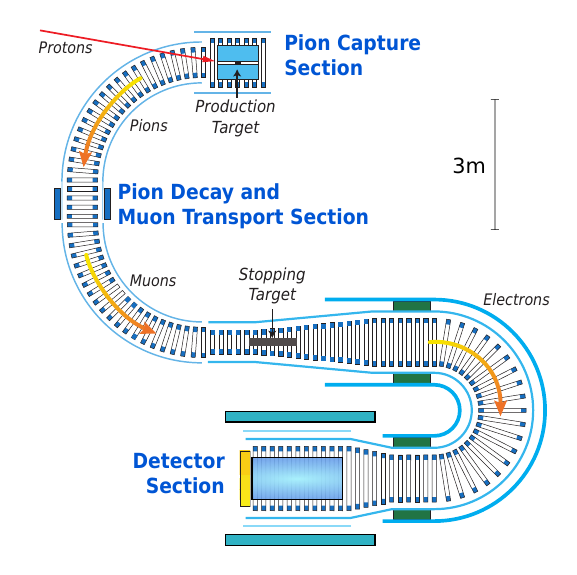}\label{fig:COMETIIdet}}
  \caption{Schematic of the COMET experiment in phase~I and
    II~\cite{Krikler:2015msn,COMET:2018auw}. }
  \label{fig:COMETdet}
\end{figure}

The second upcoming experiment at J-PARC is the two-staged COherent Muon to
Electron Transition experiment.
COMET searches for $\mu$-to-$e$ conversion on Aluminum.
A schematic of the experiment in its two stages is shown in
Fig.~\ref{fig:COMETdet}. \\
In contrast to DeeMe, the muon production and muon stopping target are
separated which allows for a more efficient suppression of beam-induced
background.
Muons are produced by directing a pulsed proton beam with bunch intervals of
around \SI{1.17}{\micro\second} on a graphite target.
In phase~I, the stopping target is separated from the production target by a
\SI{90}{\degree} transport solenoid (see Fig.~\ref{fig:COMETIdet}).
The stopping target is built from aluminum discs located inside the
detector section and delivers stopping rates of \SI{1.2e9}{\muons\per\second}.   
It is surrounded by the Cylindrical Detector (CyDet) consisting of a
cylindrical drift chamber and a trigger hodoscope built from two layers of
plastic scintillators.
The additional StrECAL detector---a combination of a straw tube tracker and an
electromagnetic calorimeter built from LYSO crystals read out by avalanche
photodiodes---performs direct beam measurements.
The StrECAL is moreover a prototype for the phase~II detector.
The Cosmic Ray Veto is built from scintillators as well as glass Resistive
Plate Chambers.\\
The COMET phase~I experiment is currently under construction and is planned to
commence data taking in 2024/2025. 
A sensitivity of $\R_{\mu e}(\mu^-\text{Al}\to
e^-\text{Al})\lesssim\num{7.0e-15}$ at \CL is expected.

In phase~II, the production target will be replaced by a tungsten target.
Combined with improvements on the proton beam and capture solenoid, a muon
production increased by a factor of twenty is expected.
The muon transport solenoid will be extended to a full \SI{180}{\degree}
turn.
The muon stopping target will be separated from the detector section by a
C-shaped electron spectrometer and the muon stopping efficiency will be
increased. 
The COMET~II experiment will be sensitive to $\R_{\mu e}(\mu^-\text{Al}\to
e^-\text{Al})<\num{3.2e-17}$ at \CL.

As a successor of COMET, there is another upgrade under investigation at
J-PARC, the Phase Rotated Intense Slow Muon source PRISM which features a muon
storage ring and the PRIsm Muon to Electron conversion experiment
PRIME~\cite{KUNO2005376}.
The PRISM/PRIME project is expected to reach sensitivities on $\R_{\mu e}$ of
$\num{e-18}$. 

\begin{figure}
  \centering
  \includegraphics[width=\textwidth]{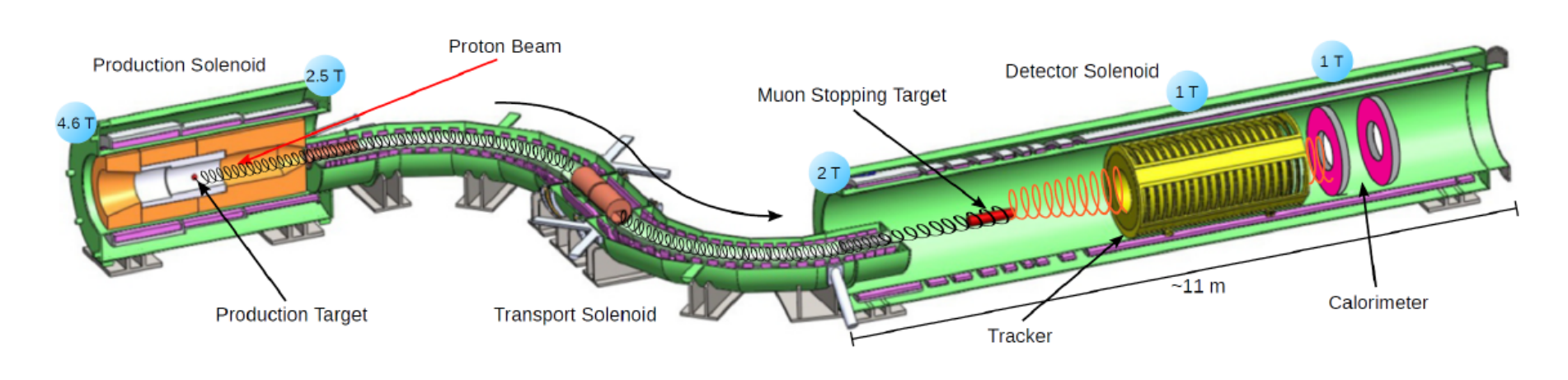}
  \caption{Schematic of the Mu2e experiment~\cite{Mu2e:2014fns}. }
  \label{fig:Mu2edet}
\end{figure}

The Mu2e experiment at Fermilab also searches for $\mu^-\text{Al}\to
e^-\text{Al}$ in two stages~\cite{Mu2e:2014fns}.
A schematic of the experiment is shown in Fig.~\ref{fig:Mu2edet}.\\
A pulsed \SI{8}{\giga\eV} proton beam hits a tungsten muon production target
with a pulse every \SI{1.7}{\micro\second}.
The production target is separated from the muon stopping target and detector
section by an S-shaped transport solenoid.
The stopping target is located in front of the tracker and calorimeter inside
the detector solenoid.
In phase~I, muon stopping rates of \SI{e10}{\muons\per\second} are envisaged.
The tracking detector is a straw tube tracker while the calorimeter is
constructed from scintillating crystals with SiPM readout.
Each side of the detector solenoid hall is covered with four layers of
scintillators which serve as cosmic ray veto. \\
The phase~I experiment is currently under construction and is expected to
commence data taking in 2025. 
It will have a sensitivity of $\R_{\mu e}(\mu^-\text{Al}\to
e^-\text{Al})<\num{6.2e-16}$ at \CL.

The Mu2e phase~II experiment will be operated at a proton beam intensity
increased by a factor of ten enabled by the Proton-Improvement-Plan-II PIP-II
at Fermilab.
Combined with improvements on all systems, sensitivities of $\R_{\mu
  e}(\mu^-\text{Al}\to e^-\text{Al})<\num{6e-17}$ at \CL are in reach of Mu2e
phase~II.

\section{Complementary of Muon cLFV Searches}

\begin{figure}
  \centering
  \subfloat[Constraints on the Wilson coefficients of an effective dipole
  $\left(C_\text{L}^\text{D}\right)$ vs.\,vector
  $\left(C_{ee}^\text{VRR}\right)$ interaction. ]{\includegraphics[width=0.3\textwidth]{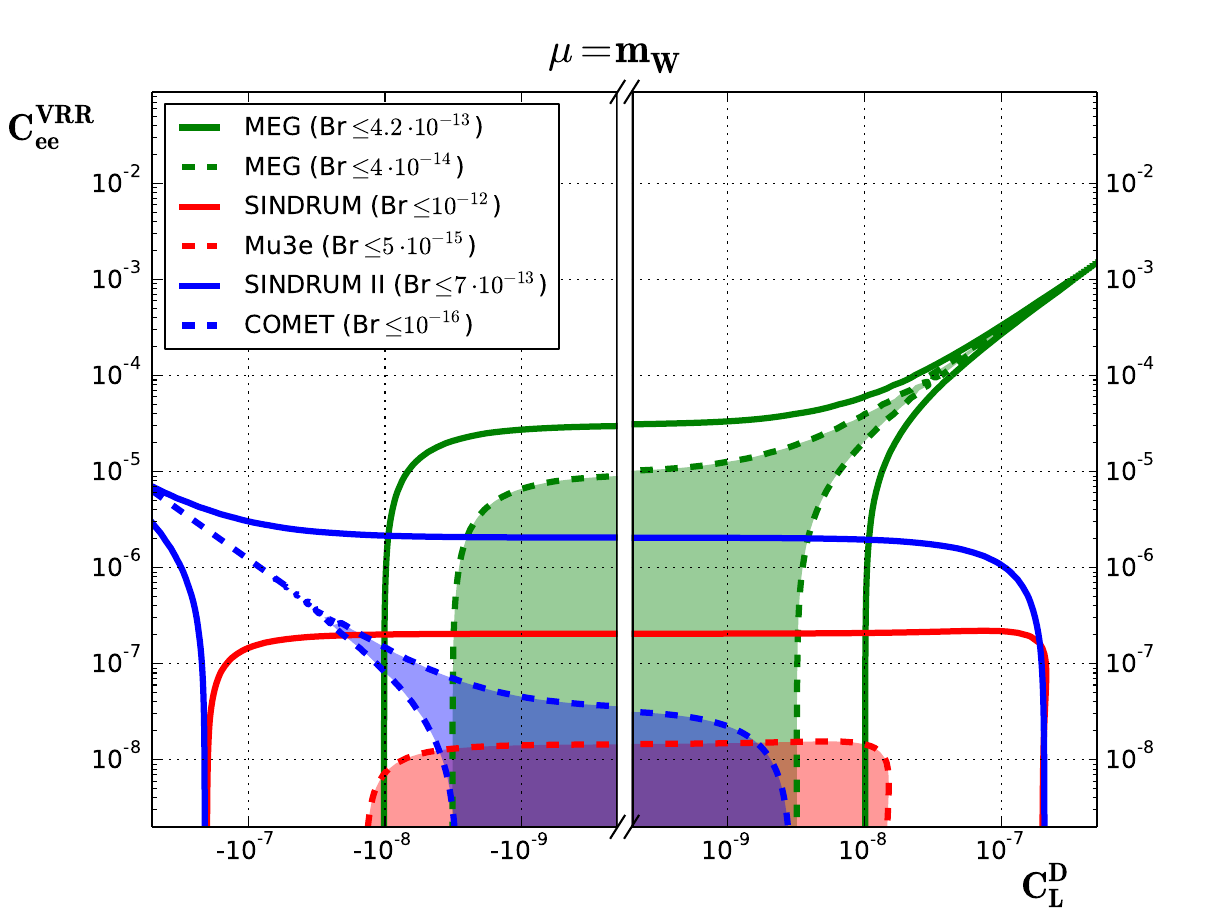}}\hfill
  \subfloat[Constraints on the Wilson coefficients of an effective scalar
  $\left(C_{ee}^\text{SLL}\right)$ vs.\,vector
  $\left(C_{ee}^\text{VRR}\right)$ interaction. ]{\includegraphics[width=0.3\textwidth]{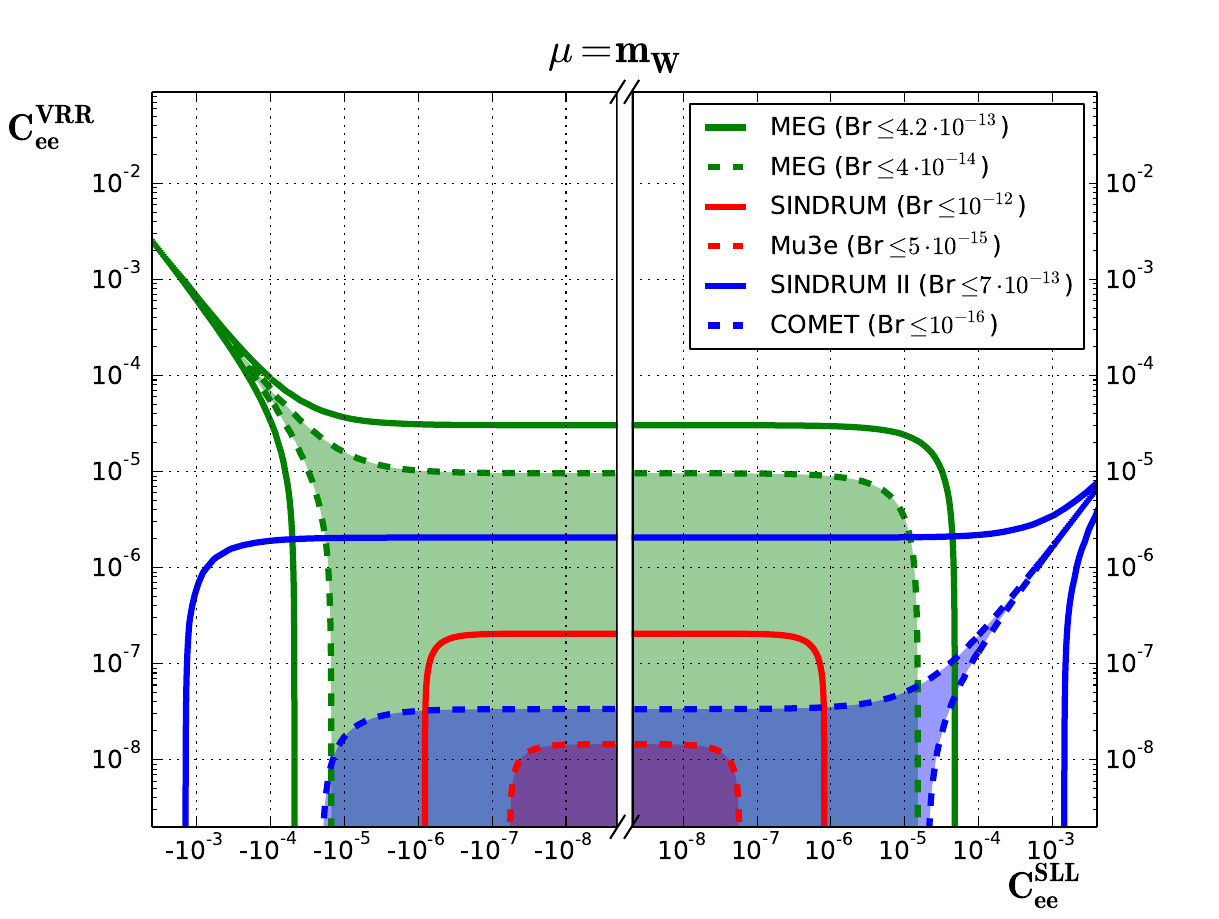}}\hfill
  \subfloat[Constraints on the Wilson coefficients of an effective dipole
  $\left(C_\text{L}^\text{D}\right)$ vs.\,scalar
  $\left(C_{bb}^\text{SLR}\right)$ interaction. ]{\includegraphics[width=0.3\textwidth]{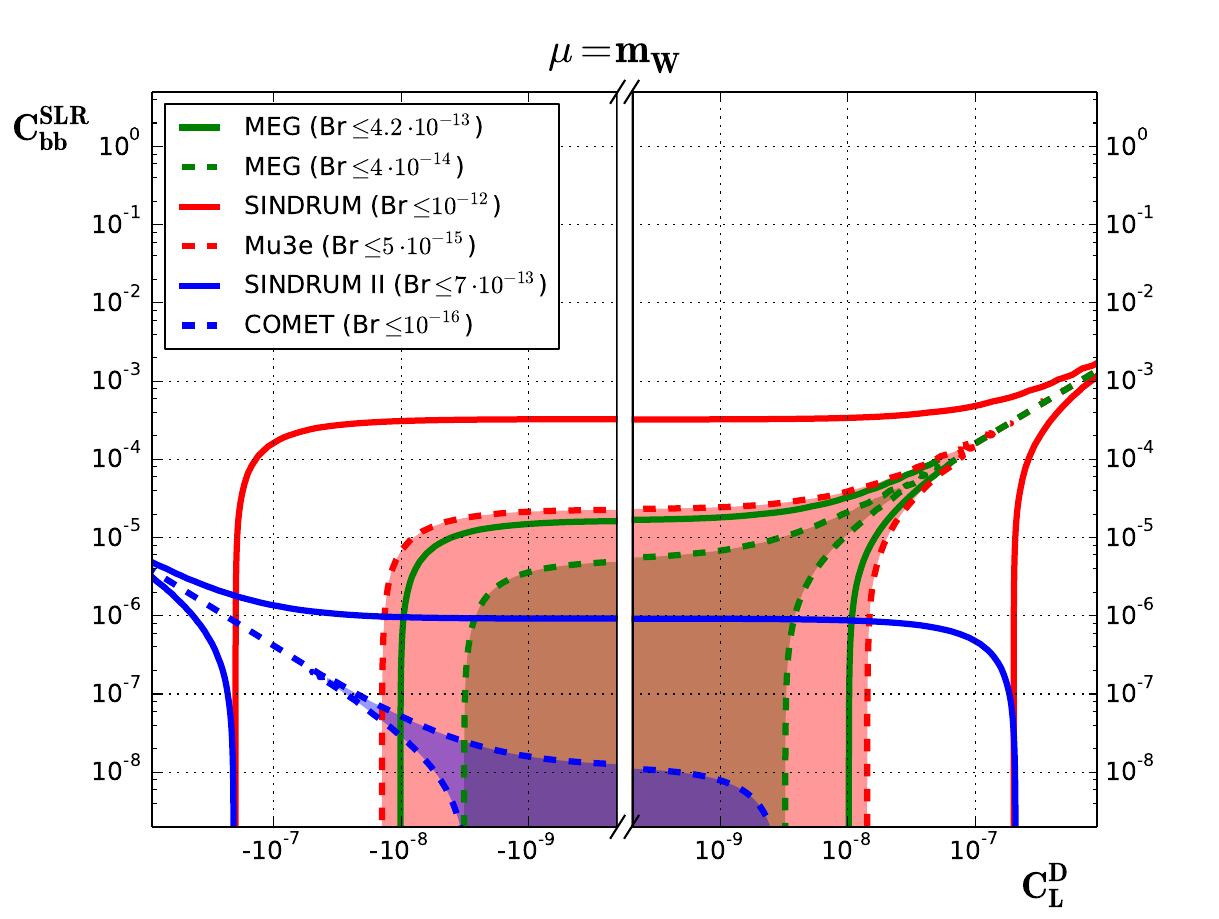}}
  \caption{Observed and prospected constraints on Wilson coefficients in muon
    LFV effective field theories from current and future searches for
    \muegammacharged, \mueeecharged and
    \muNeNcharged~\cite{Crivellin:2017rmk}. }
  \label{fig:cLFVeff}
\end{figure}

The sensitivity of the three golden channels to different types of BSM
interactions can be studied and compared by means of effective field theories
(see for example~\cite{Crivellin:2017rmk}).
As shown in Fig.~\ref{fig:cLFVeff}, each channel has specific strengths and
weaknesses in constraining the various interactions.
Thus, the interplay of potential observations and non-observations in the muon
LFV channels will allow to draw a more complete picture of favoured and
disfavoured BSM models.

\begin{figure}
  \centering
  \includegraphics[width=0.45\textwidth]{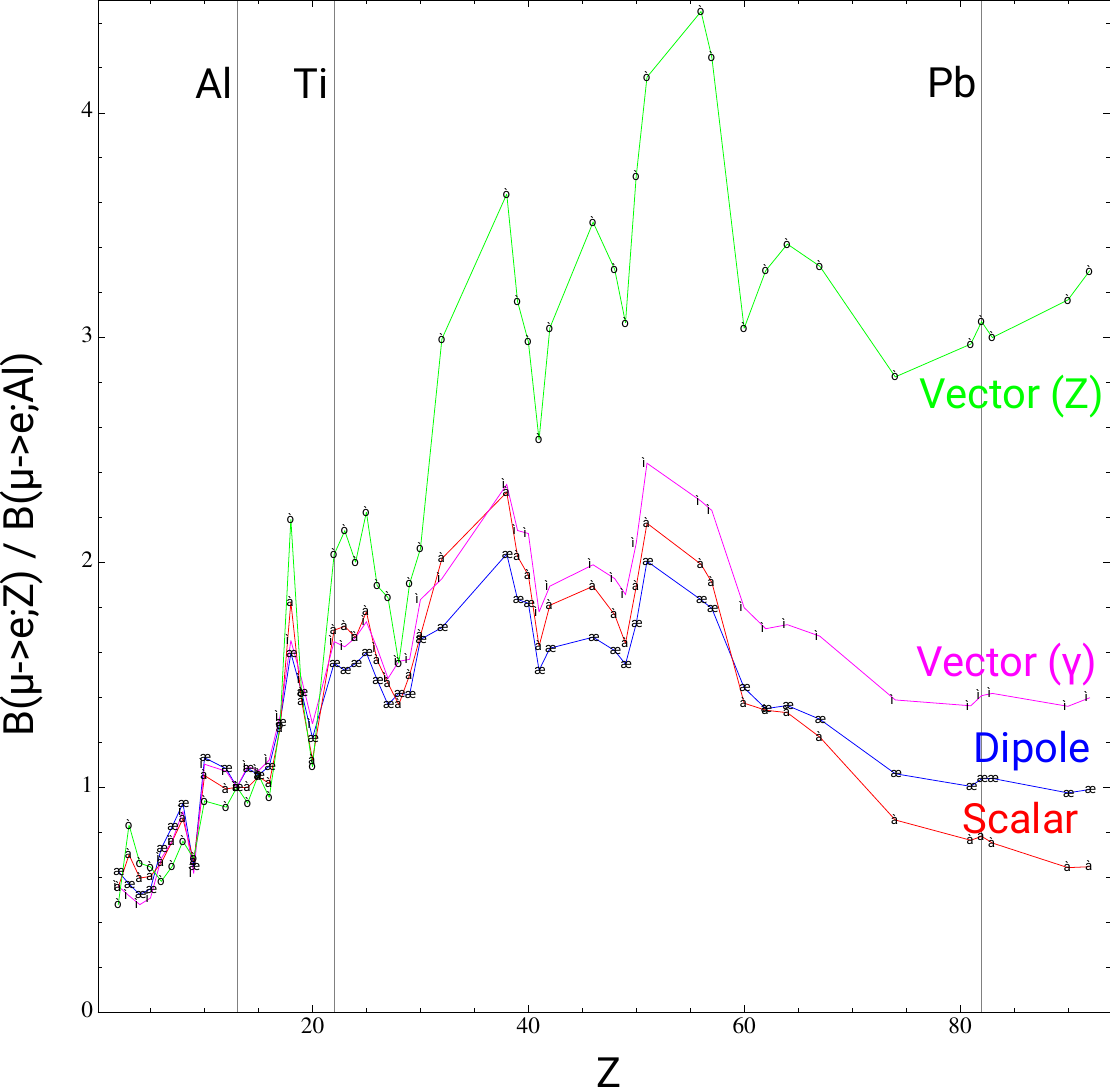}
  \caption{Dependence on the conversion rate in \muNeNcharged transitions
    mediated via different effective interactions on the atomic number $Z$ of
    the nuclei. Adapted from~\cite{Cirigliano:2009bz}. }
  \label{fig:muNeNtarget}
\end{figure}

\begin{figure}
  \centering
  \subfloat[Effective dipole operator. ]{\includegraphics[width=0.3\textwidth]{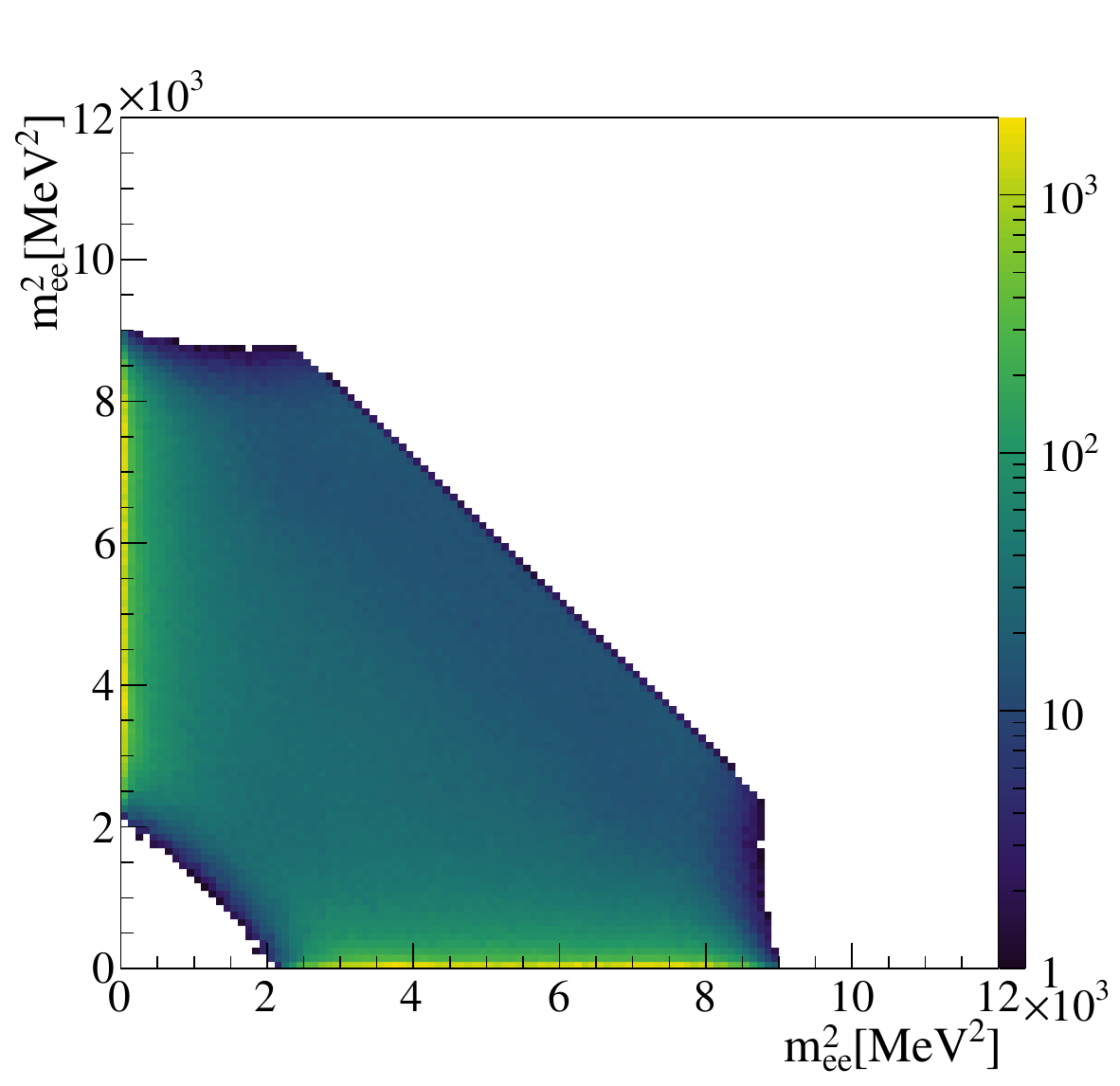}}\hfill
  \subfloat[Effective vector operator. ]{\includegraphics[width=0.3\textwidth]{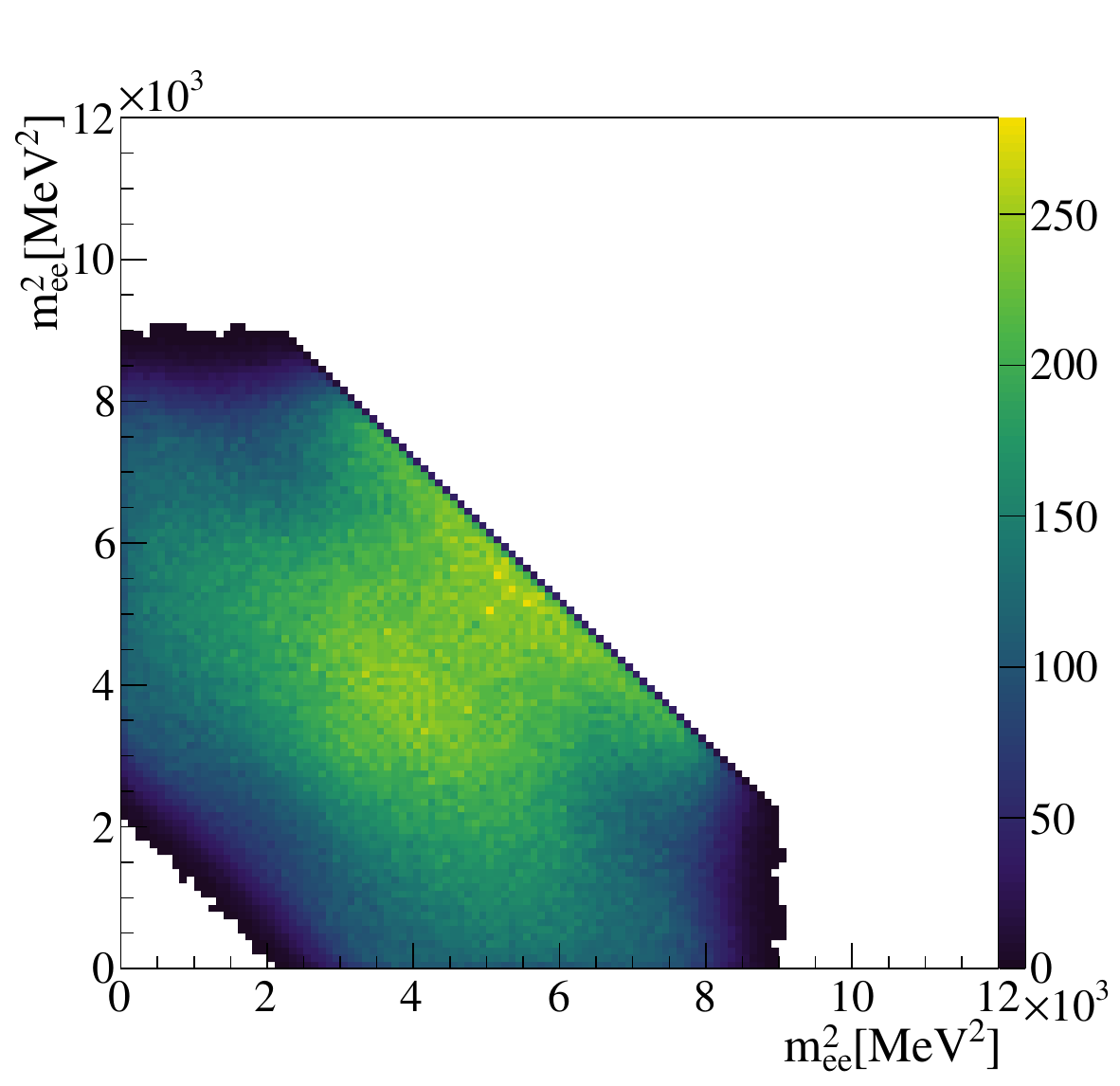}}\hfill
  \subfloat[Effective scalar operator. ]{\includegraphics[width=0.3\textwidth]{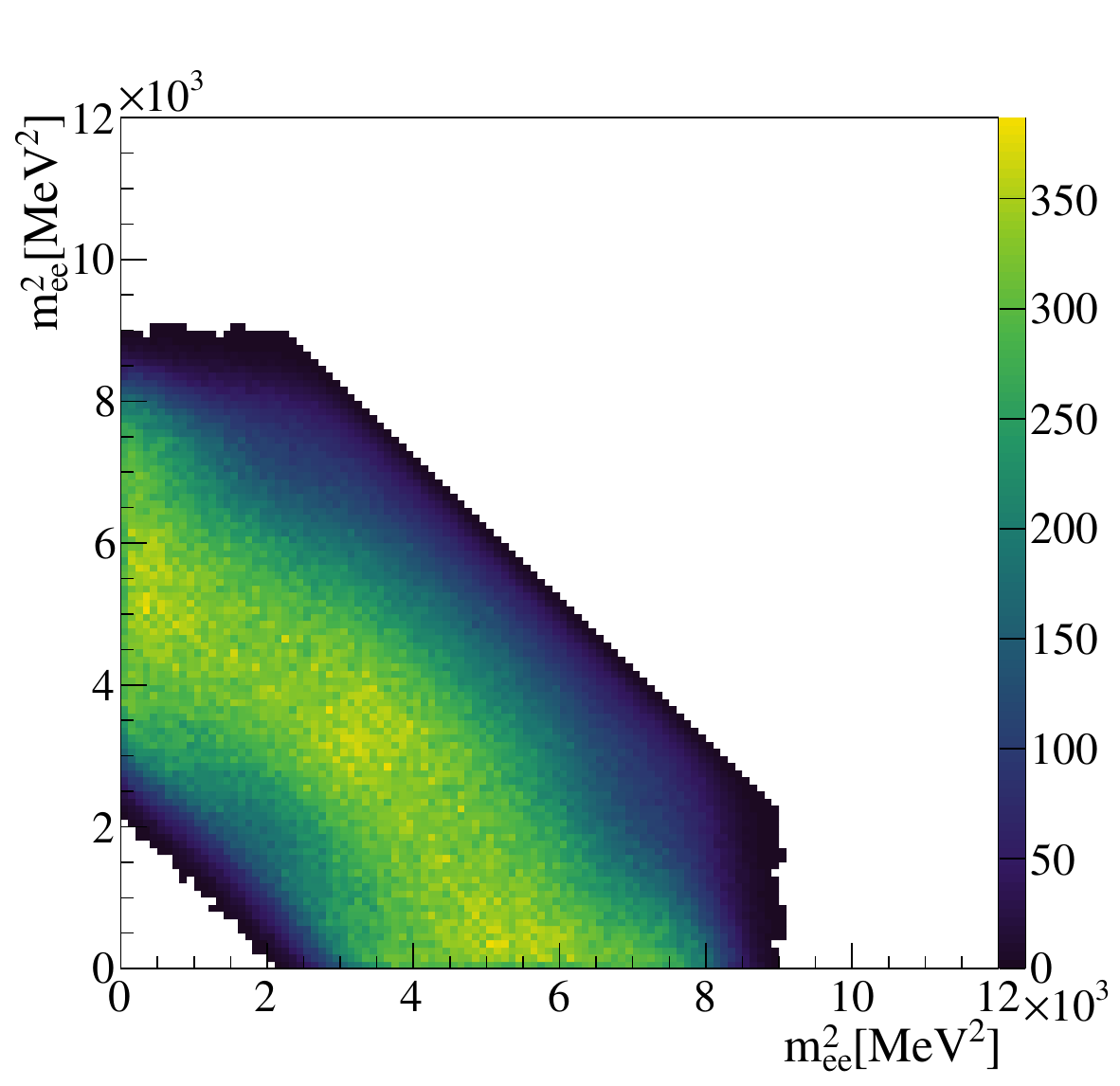}}
  \caption{Dalitz plots of the invariant mass squared of $e^+e^-$ pairs from
    \mueeecharged mediated via various effective operators in the Mu3e phase~I
    experiment~\cite{Perrevoort:2018ttp}.
    The effective Lagrangian is taken from~\cite{Kuno:1999jp}. }
  \label{fig:Mu3eeff}
\end{figure}

Furthermore, it is possible to distinguish certain effective operators
directly in \muNeNcharged and \mueeecharged searches. \\
In \muNeNcharged, the dependence of the conversion rate on the atomic number
$Z$ of the nucleus varies for the different operators (see
Fig.~\ref{fig:muNeNtarget}). 
Thus, searches with different target materials can be used to narrow down the
type of BSM interaction.
However, the choice of suited target materials is limited as the lifetime of
the muonic atom needs to be sufficiently long and as well as for technical
reasons.
In addition to aluminum, for example titanium, vanadium and lithium are
considered. \\
In the case of \mueeecharged, the kinematics of the final decay products can
reveal the type of BSM interaction.
In Fig.~\ref{fig:Mu3eeff}, Dalitz plots of the invariant mass squared of the
two possible $e^+e^-$ pairs are shown.
Different effective operators reveal characteristic distributions that allow
to distinguish various BSM interactions.


\section{Exotic Physics at  Muon Experiments}

Despite being designed to investigate one specific channel, the above
mentioned experiments are not limited to a single measurement.
A selection of searches that go beyond the scope are presented in the following. 

The ATOMKI collaboration has reported the observation of an excess in the
angular distributions of internal pair creation in the transition
$^7\text{Li}(p, e^+e^-)^8\text{Be}$~\cite{Krasznahorkay:2015iga}
as well as in the transition
$^3\text{H}(p,e^+e^-)^4\text{He}$~\cite{Krasznahorkay:2020wwp}. 
This excess would be compatible with the production and subsequent decay of a
hypothetical BSM boson at a mass of around \SI{17}{\MeV} commonly referred to
as $X(17)$. \\
The MEG~II experiment will repeat this measurement using a $p$ beam from a
Cockcroft-Walton accelerator and a $\text{Li}_2\text{B}_4\text{O}_7$ target
which are normally used for calibration purposes~\cite{Alves:2023ree}.
The collaboration has taken first data for this measurement. \\
In addition, the Mu3e collaboration will perform searches for $e^+e^-$
resonances in \mueeenunucharged in view of searches for dark
photons~\cite{Perrevoort:2018ttp}. 
This search will also be sensitive to $X(17)$.

\begin{figure}
  \centering
  \includegraphics[width=0.6\textwidth]{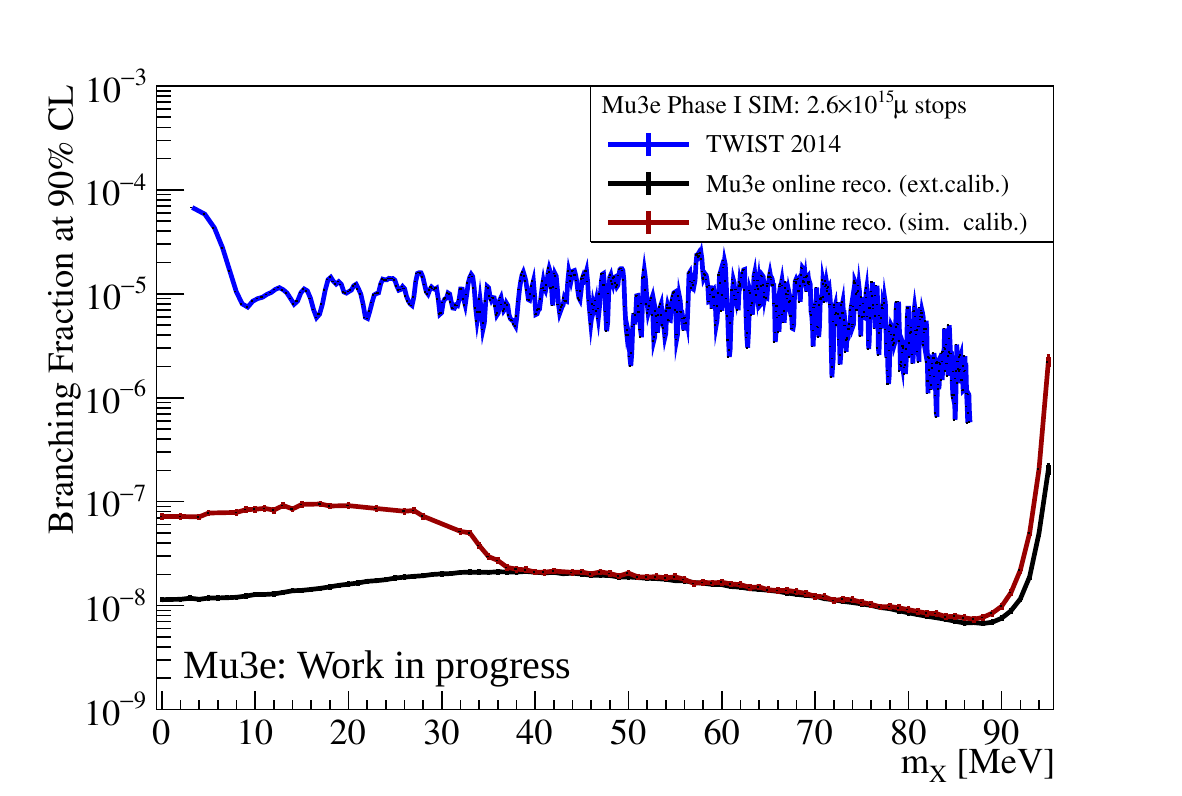}
  \caption{Sensitivity of the Mu3e phase~I experiment to \mueXcharged compared
    to current most stringent limits set by TWIST~\cite{Perrevoort:2018ttp}.
    Two scenarios for the calibration of the momentum scale are shown: one
    assuming calibration using an external process (\emph{ext.\,calib.}),
    and the other one assuming simultaneous calibration using the momentum
    spectrum of \muenunucharged decays (\emph{sim.\,calib.}).
    TWIST results by courtesy of R.\,Bayes~\cite{TWIST:2014ymv}.}
  \label{fig:mueX}
\end{figure}

Another channel that will be investigated with the Mu3e experiment is
\mueXcharged in which $X$ is an axion-like particle from a broken flavour
symmetry like a familon or majoron~\cite{Wilczek:1982rv, Calibbi:2020jvd} that
leaves the detector unseen.
For this purpose, the Mu3e data acquisition is adapted to accommodate online
histogramming of track fit results such as momenta and emission angles of
events with single positrons on the event filter farm.
Current limits on \mueXcharged are set by Jodidio et al.\,at $\BR(\mu^+\to
e^+X)<\num{2.6e-6}$ at \CL for massless $X$~\cite{Jodidio:1986mz}, and by the
TWIST collaboration at $\BR(\mu^+\to e^+X)<\num{9e-6}$ at \CL on average for
$X$ with masses between \SI{13}{\MeV} and \SI{80}{\MeV}.
The sensitivity of the Mu3e experiment in phase~I exceeds the limits set by
TWIST by two orders of magnitude in a large range of $X$ masses (see
Fig.~\ref{fig:mueX}) and will be further improved in phase~II due to a twenty
times larger number of observed muon decays and an enhanced detector
performance. \\
Searches for \mueXcharged will be also performed at
MEG~II~\cite{Calibbi:2020jvd, Jho:2022snj}
and at the \muNeN experiments Mu2e and COMET~\cite{BoundmueaSnowmass,
  Xing:2022rob}.
MEG~II will further investigate the channel \muegammaXcharged.

Furthermore, the lepton flavour and lepton number violating process $\mu^-
N\to e^+N^\prime$ can be investigated at the muon conversion experiments Mu2e
and COMET~\cite{Lee:2021hnx}, however they require a better theoretical
understanding of the background contribution of radiative muon capture
$\mu^- N\to \nu_\mu N^\prime\gamma$ with $\gamma\to e^+e^-$.
Current upper limits are set by SINDRUM~II measurements with a titanium target
at $\R(\mu\text{Ti}\to e^+\text{Ca})<\num{1.7e-12}$ at
\CL~\cite{SINDRUMII:1998mwd}.



\section{Summary}

The observation of cLFV $\mu$-to-$e$ transitions would be an unambiguous
sign of BSM physics.
Several experiments are currently operating or under construction which will
investigate these transitions with an up to four orders improved sensitivity
compared to current limits:
the ongoing MEG~II experiment at PSI searching for \muegammacharged, the
upcoming Mu3e experiment at PSI searching for \mueeecharged, and the ongoing
DeeMe experiment and the upcoming COMET experiment at J-PARC as well as the
upcoming Mu2e experiment at Fermilab searching for $\mu$-to-$e$ conversion
on nuclei \muNeNcharged. 
The complementarity of these searches allows to narrow down the type of BSM
interaction in case of discovery or strongly constrain numerous BSM models in
case of non-observation. 
In addition, further BSM signatures can be investigated by these experiments
with competitive sensitivity.

\bibliographystyle{JHEP}
\bibliography{myBib}

\acknowledgments
The author's work is funded by the Federal Ministry of Education and Research
(BMBF) and the Baden-W\"urttemberg Ministry of Science as part of the
Excellence Strategy of the German Federal and State Governments.

\end{document}